\documentstyle[twoside,fleqn,espcrc2,epsfig,here]{article}

\newcommand{\AmS}{{\protect\the\textfont2
  A\kern-.1667em\lower.5ex\hbox{M}\kern-.125emS}}

\title{Future High Energy Neutrino Telescopes}

\author{C.Spiering\address {DESY Zeuthen \\ 
        D-15738 Zeuthen, Germany}}

\begin{document}

\begin{abstract}
This talk summarizes the main physics goals and basic methods 
of telescopes for high energy neutrinos. It reviews the present
status of deep underwater telescopes and sketches the ICECUBE
project as an example for a cube kilometer detector.
It is suggested to develop techniques for radio and acoustic
detection hand in hand with big optical arrays. These large arrays
should be complemented by medium-size detectors in the Megaton range.
\end{abstract}

\maketitle
\vspace{-1mm}

\section{Physics and Methods}

Whereas MeV neutrino astronomy has been established by the observation 
of solar 
neutrinos and neutrinos from supernova SN1987, 
neutrinos with energies of GeV to PeV 
which must accompany the production of  high 
energy cosmic rays still await discovery. 
Detectors underground have turned out to be too 
small to detect the feeble fluxes of 
energetic neutrinos from cosmic accelerators. 
The high energy frontier is being tackled by 
much larger, expandable arrays constructed in open water or ice. 
The physics goals of high energy neutrino telescopes have been 
covered in detail by 
other talks of this conferences. They include:
\vspace{-1mm}
\begin{itemize}
\item[a)] Search for neutrinos from cosmic acceleration processes in 
  galactic sources like
 binary  systems or supernova remnants (SNR), or extragalactic 
sources like active galactic 
nuclei (AGN) or gamma ray bursters (GRB) \cite{Ghandi,Waxman}.
\vspace{-1mm}
\item[b)] Search for ultra high energy neutrinos 
from topological defects (TD) \cite{Weiler}.
\vspace{-1mm}
\item[c)] Search for neutrinos from the annihilation of Weakly 
Interacting Massive Particles 
(WIMPs) \cite{Gondolo}.
\vspace{-1mm}
\item[d)] Search for magnetic monopoles.
\vspace{-1mm}
\item[e)] Investigation of neutrino oscillations, using neutrinos 
from accelerators, 
the atmosphere or extraterrestrial neutrinos \cite{oscill}.
\vspace{-1mm}
\item[f)] Monitoring for MeV neutrinos from supernova bursts 
in our Galaxy \cite{Scholberg}.
\vspace{-1mm}
\item[g)] Cosmic ray physics with atmospheric muons
\end{itemize}
\vspace{-1mm}
The telescopes presently under construction detect the Cherenkov 
light generated by 
secondary particles produced in neutrino interactions. They are 
optimized for the detection 
of muon {\it tracks} and for 
energies of a TeV or above, by the following reasons: 
\vspace{-1mm}
\begin{itemize}
\item[a)] The flux of neutrinos from cosmic accelerators is 
expected to behave  like $E_{\nu}^{-2.0 \div 2.5}$ 
whereas the spectrum of atmospheric neutrinos above 
100\,GeV falls like $E_{\nu} ^{-3.7}$,
yielding a better signal-to-background ratio at higher energies.
\vspace{-1mm}
\item[b)] Neutrino cross section and muon range increase with energy. 
Due to the large muon 
range, the effective volume of the detectors may considerably 
exceed their geometrical volume.
\vspace{-1mm}
\item[c)] The mean angle between muon and neutrino decreases with energy like
$ E^{-2}$, with a 
pointing accuracy of about 1.5$^o$ at 1\,TeV.
\vspace{-1mm}
\item[d)] Mainly due to pair production and bremsstrahlung, the energy loss of 
muons increases 
with energy. For energies above 1\,TeV, 
this allows to estimate the muon energy from 
the larger light emission along the track.
\end{itemize}
 
There are questions which require a threshold in the range a several 
GeV or  a few tens of 
GeV, like the study of neutrino oscillations or the search 
for neutrinos from WIMP 
annihilation. Since it is hard to combine large detection area  
at high energies (which 
suggests a large spacing of detector elements) 
with a low energy threshold (which require 
a small spacing), one may consider the worldwide operation 
of several complementary 
detectors.

Apart from elongated tracks, {\it cascades} can be detected. 
Their length increases only 
like the logarithm of the cascade energy. With
typically 5-10 meters length, 
cascades may be considered 
as quasi point-like compared 
to the spacing of photomultipliers in Cherenkov telescopes. 
The effective volume  
for cascade detection is close to the geometrical volume. While
for present telescopes it therefore is much smaller than that for
muon detection, for kilometer-scale detectors and not
too large energies it can reach the same order of magnitude
like the latter.

Ultra-high energy cascades could be detected not only by their
Cherenkov light but also by coherent Cherenkov waves in the radio 
range or by acoustic pulses. With an attenuation length on the kilometer
scale, these techniques allow detection volumes of Giga-tons or
even higher. Since the initial energy has to be very high
to yield a detectable signal at all, they may start to compete with
optical detectors only above a few PeV. 

Figure\,1 sketches the detector masses and energy ranges which 
are characteristic for underground detectors, for optical
detectors in water or ice, and for acoustic and radio
detectors.

\vspace{-2mm}

\begin{figure}[htb]
\vspace{9pt}
\centering
\mbox{\epsfig{file=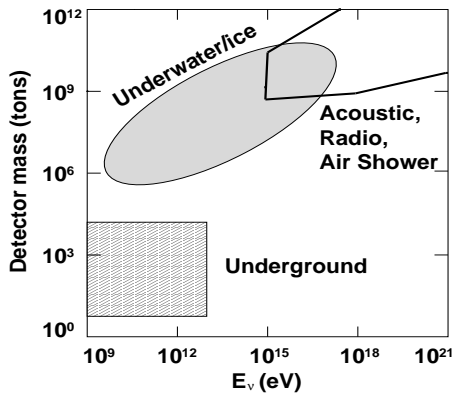,height=5.1cm,width=6.0cm}}
\vspace{-0.8cm}
\caption[2]{\small
Domains of different detection techniques. 
}
\end{figure}



\section{Optical Detectors under water and ice}

Optical underwater detectors consist of a lattice of 
photomultipliers (PMs) spread over a large open volume  
in the ocean, in lakes or in 
ice. The PMs record arrival time and amplitude of Cherenkov light
emitted by muons or cascades.

The development in this field was largely stimulated 
by the DUMAND project \cite{Dumand} which 
was cancelled in 1995. The other pioneering experiment is the 
Baikal telescope \cite{Domo}.
The Baikal collaboration not only was the first to deploy 
three strings 
(as necessary for full spatial reconstruction \cite{Bai1}), 
but also reported the first atmospheric neutrinos
detected unterwater \cite{Bai2}. At present, NT-200, 
an array comprising 192 PMs, is 
taking data. With respect to its size, NT-200 has been surpassed by 
the AMANDA detector at the 
South Pole \cite{Ama1,Barwick}.  
With 677 PMs, the present AMANDA-II array reaches an area of 
a few $10^4$\,m$^2$ for 1 TeV muons. Although still far below the square
kilometer size suggested by most theoretical models 
\cite{Ghandi,Weiler,GHS}, AMANDA-II may be the first detector with a realistic 
discovery potential with respect to extraterrestrial high energy 
neutrinos. Limits obtained from the analysis of data taken with the 
three times smaller AMANDA-B10 in 1997 have been reported at this 
conference \cite{Barwick}. The limit  on the diffuse flux from unresolved 
sources with an assumed $E^{-2}$ spectrum is of the order of 
$10^{-6} \, E_\nu \, ^{-2}$\,cm$^{-2}$\,s$^{-1}$\,sr$^{-1}$\,GeV$^{-1}$, 
close to the recent bound established by Mannheim, Protheroe and Rachen
\cite{MPR} but still above the bound given by Waxmann and Bahcall 
\cite{WB}. The flux limit on point sources with an  $E^{-2}$  
 spectrum at the northern sky (declination larger than 30 degrees) 
is of the order of $10^{-7}$\,cm$^{-2}$\,s$^{-1}$ for $E_{\nu}  > 10$\,GeV.
 The overall sensitivity of AMANDA-B10 has been verified by a  
sample of more than 200 events which is dominated by atmospheric neutrinos.

Two projects for large neutrino telescopes are underway in the 
Mediterranean - ANTARES \cite{Antares} and NESTOR \cite{Nestor}. 
Both have 
assessed the relevant physical and optical parameters of their sites, 
developed deployment methods and performed a series of operations 
with a few PMs. ANTARES and NESTOR envision different deployment 
schemes and array designs. The NESTOR group plans to deploy a tower 
of several floors, each carrying 14 PMTs. The ANTARES detector will 
consist of 13 strings, each equipped with 30 storeys and 3 PMTs per 
storey. This detector will have an area of about $3 \cdot  10^4$\,m$^2$
 for 1\,TeV
 muons - similar to AMANDA-II - and is planned to be fully deployed 
by the end of 2003. On top of these two advanced projects, there is 
an Italian initiative, NEMO \cite{NEMO}, which studies a appropriate 
sites for a future km$^3$ detector and started with $R\&D$ activities.

There have been longstanding discussions about the best location for 
future km$^3$ telescopes. What concerns geographic location, 
one detector on each hemisphere would be ideal for full sky coverage. 
With respect to optical properties, water detectors 
in oceans seem to be favored: 
although the absorption length  of  Antarctic ice at Amanda depths 
is nearly twice as long as in oceans (and about four times that of 
Baikal), ice is characterized by strong light scattering, and its 
optical parameters vary with depth. Light scattering leads
 to a considerable delay of Cherenkov photons.
On the other hand 
ice does not suffer from the high potassium content of ocean water or 
from bioluminescence. These external light sources result in counting
 rates ranging from several tens of kHz to a few hundred kHz, compared
 to less than 500 Hz pure PM dark count rate in ice.  Depth arguments 
favor oceans.  Note, however, that this argument lost its initial 
strength after BAIKAL and AMANDA had developed reconstruction methods  
which effectively reject even the high background at shallow depths. 
What counts most, at the end, are basic technical questions like 
deployment, or the reliability of the single components as well as 
of the whole system. Systems with a non-hierarchical structure like 
AMANDA (where each PM has its own  2\,km cable to surface) will  
suffer less from single point failures than water detectors do.  
In the case of water,  longer distances between detector and shore 
station have to be bridged. Consequently, not every PM can get its 
own cable to shore, resulting in a strictly hierarchical system 
architecture. This drawback of water detectors may be balanced by 
the fact that they allow retrieval and replacements of failed components
 - as the BAIKAL group has demonstrated over many years.

Most likely, the present efforts will converge to two km$^3$ detectors 
for very high energy neutrinos, one at the South Pole and one in the 
Mediterranean (with the Baikal site possibly filling a niche a 
intermediate energies). Representatively for these large telescopes, 
the following section sketches ICECUBE, the most advanced design 
for such an detector.

\section{ICECUBE}

The proposed ICECUBE detector consists of 4800 PMTs deployed on 80 
vertical strings.  It is planned to install up to 16 strings per 
austral summer season. 
Fig. 2 gives a top view of the proposed 
geometry and the position of the array with respect to  AMANDA-II
and the air shower array SPASE-2.

\vspace{-0.6cm}
\begin{figure}[htb]
\vspace{9pt}
\centering
\mbox{\epsfig{file=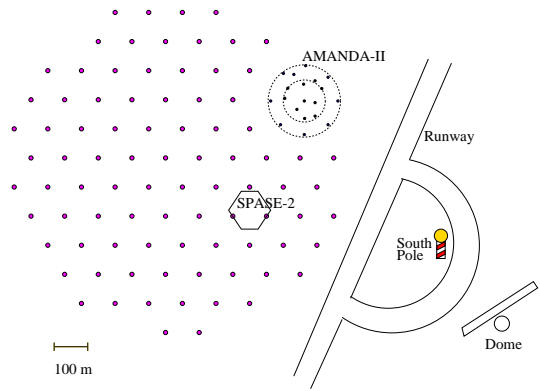,width=7cm}}
\vspace{-0.4cm}
\caption[2]{\small
Top view of the proposed ICECUBE detector
}
\end{figure}
\vspace{-0.5cm}

The detailed geometry is still being optimized. The default geometry 
assumes a string spacing of  125 m and distances of  PMTs along a 
string of 16 m.  In the course of MC simulations, these parameters
have been varied over a wide range.
Apart from uniform arrays, also nested configurations with sub-arrays 
of higher density have been simulated \cite{Matthias,Proposal}.
Fig.3 
gives the effective area as a function of string spacing. Clearly, 
detection of 100 GeV muons gets worse for the large spacing prefered 
at energies above a TeV. Raising the number of PMTs leads to a lower  
energy threshold provided the side length of the detector is kept 
constant. On the other hand, the maximum effect for high energy muons 
is obtained if not only the number of PMTs   but also the spacing is 
increased. Clearly, the final configuration is a function of physics 
priorities.  I mention a few of them in the following.

\vspace{-0.5cm}

\begin{figure}[htb]
\vspace{9pt}
\centering
\mbox{\epsfig{file=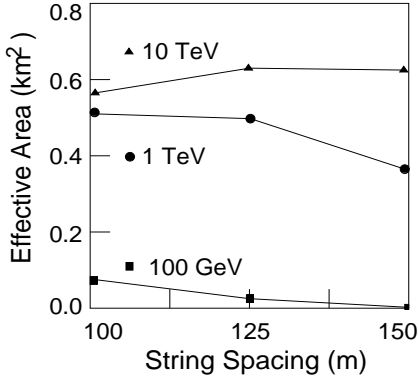,height=5.0cm,width=5.5cm}}
\vspace{-0.7cm}
\caption[2]{\small
ICECUBE effective area at different energies as a function 
of string spacing
}
\end{figure}

\vspace{-1mm}
\noindent
{\it Search for extraterrestrial high energy neutrinos}
\medskip

ICECUBE would record about 35,000 (3,000) atmospheric neutrinos 
with energies of 1-10 (10-100) TeV. The number of reconstructable 
events below 1 TeV also reaches a few $10^4$ but depends strongly on the 
efficiency of background rejection. An extraterrestrial diffuse flux 
from AGN of 
$10^{-7} \, E_{\nu}^{-2}$\,cm$^{-2}$\,s$^{-1}$\,sr$^{-1}$\,GeV$^{-1}$,
 one order of magnitude 
below the theoretical bound derived in \cite{MPR}, would result in 
600 (800) events with energies of 1-10 (10-100) TeV. 
Energy reconstruction 
is crucial to identify AGN neutrinos. With present algorithms, a 
resolution $\sigma (\ln{E_{\mu}}) \approx 0.4$ has been obtained 
(i.e. slightly better than  one order of magnitude). We anticipate 
that this value can be improved  to 0.25 with the help of advanced 
methods.  If the majority 
of the AGN neutrinos would come  from a few tens of AGN,  the 
signal-to-background ratio would be improved by up to two orders 
of magnitude. This would be essential for detecting an extraterrestrial 
excess at energies below 100 TeV. 

With a pointing accuracy of about one degree or better, the easiest 
search strategy will be a point source analysis (see for details 
\cite{Barwick}). In this case, the atmospheric neutrino background is
 reduced by nearly four orders of magnitude.

Fig.4 \cite{LM} shows theoretical predictions and bounds on 
the diffuse flux of extraterrestrial 
neutrinos as well as the flux of atmospheric neutrinos. ICECUBE might
 search for a signal down to a few 
$10^{-9} \, E_{\nu} ^{-2}$\,cm$^{-2}$\,s$^{-1}$\,sr$^{-1}$\,GeV$^{-1}$.

\vspace{-5mm}
\begin{figure}[htb]
\vspace{9pt}
\hspace{-7mm}
\mbox{\epsfig{file=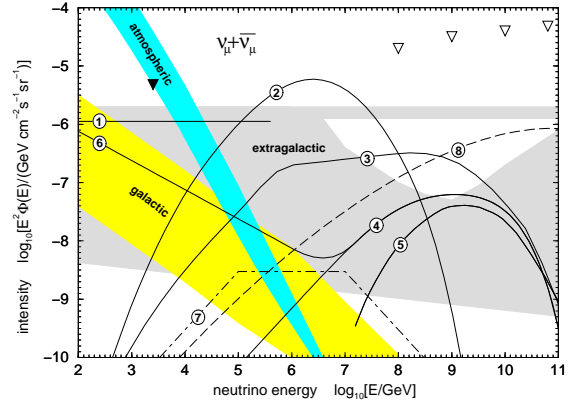,width=6cm}}
\vspace{-8mm}
\caption[2]{\small
Summary of expected $\nu_{\mu} + \overline{\nu}_{\mu}$ intensities. 
Numbered lines: 
1-4,6) AGN models (Nellen93, Stecker96, Mannheim et al., 
Mannheim 94),
5) neutrinos from interaction of UHE protons with cosmic 
microwave background (Protheroe96),
7) GRB model (Waxmann, Bahcall),
8) neutrinos from topological defects (Sigl98)
Shaded regions: theoretical bounds on diffuse fluxes from \cite{MPR}.
Bold triangle: Frejus limit on excess above atmospheric neutrinos,
blank triangle: limits on upward events from Fly's Eye.
Figure taken from \cite{LM})}
\label{abb_nu_fluxes_cosmic}
\end{figure}


\noindent
{\it Search for magnetic monopoles}
\medskip

A  magnetic monopole with unit magnetic Dirac charge $g = 137/2 \cdot e$
 and velocities above the Cherenkov threshold in water
$(\beta > 0.75)$ would emit 
Cherenkov radiation smoothly along its path, 
exceeding that of a bare relativistic muon 
by a factor of 8300 \cite{Thorn}. This is
a rather unique signature. Fig.5 summarizes 
the limits obtained until now \cite{Mon}. A cube kilometer detector could 
improve the sensitivity of this search by nearly two orders of 
magnitude. The search could be extended to even lower velocities 
by detecting the $\delta$ electrons generated along the monopole path.

\begin{figure}[htb]
\vspace{9pt}
\centering
\mbox{\epsfig{file=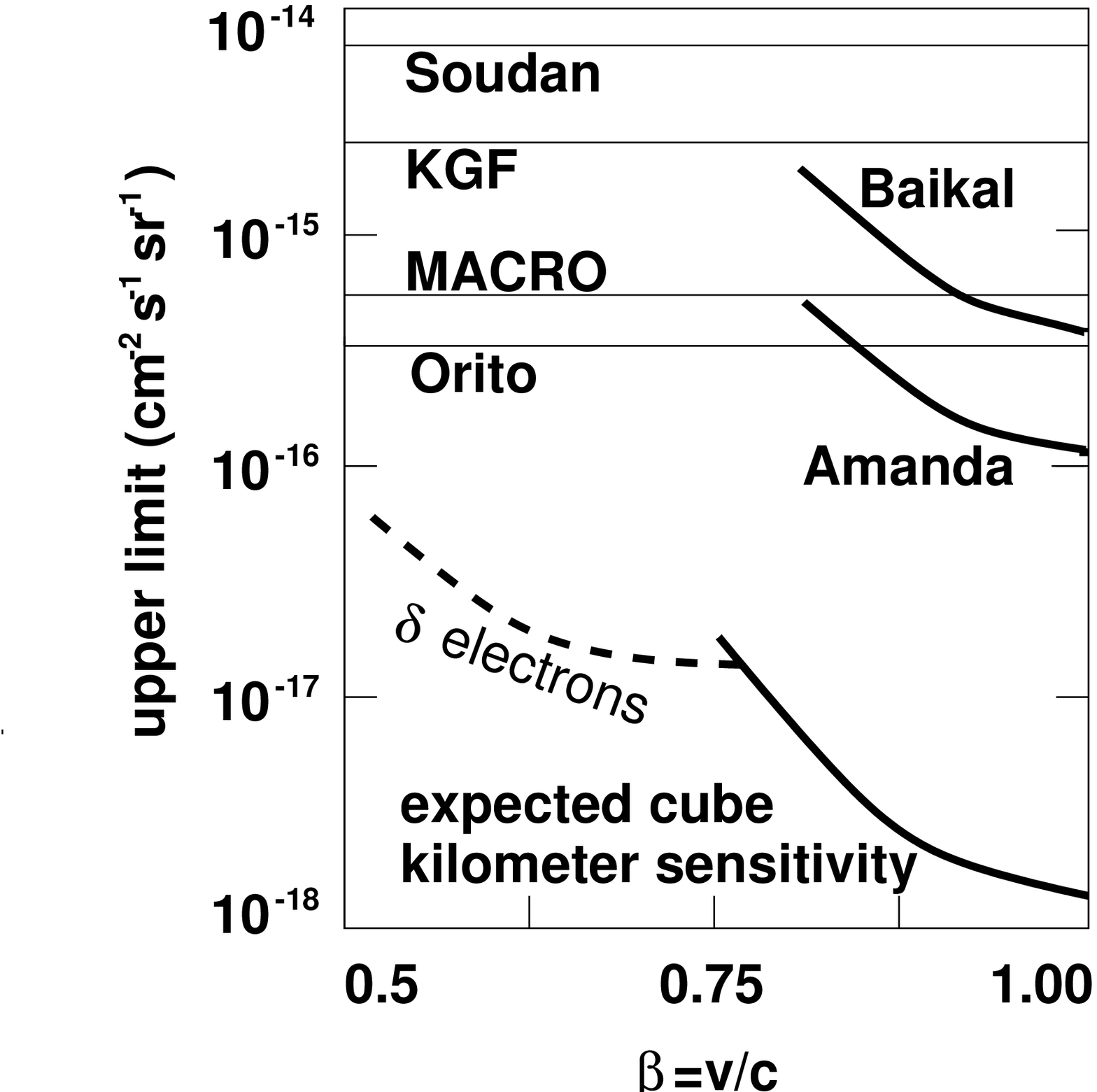,height=5.5cm,width=6cm}}
\vspace{-0.7cm}
\caption[2]{\small
Limits on the flux of relativistic monopoles
}
\end{figure}

\medskip

\noindent
{\it Indirect WIMP search}
\medskip

The annihilation of neutralinos trapped within the core of the Earth or 
in the Sun  would generate high energy neutrinos from the nearly vertical 
direction or from the Sun, respectively \cite{BEG}.  Present searches 
by experiments underground and underwater for an excess of muons from the 
center of the Earth exclude a fraction of supersymmetric models with 
neutralino masses above 50-100 GeV.  Provided a  muon detection threshold 
 of  about 20 GeV,  a cube kilometer detector might complement future 
direct search experiments  for WIMPs  like CRESST or GENIUS \cite{Edsjo}. 


\bigskip
\noindent
{\it Supernova Bursts}
\medskip

Due to the low external noise rate, already the present AMANDA array is 
sensitive to an increase of the individual counting rates of all PMs  
resulting from  a Supernova burst closer than 8-9 kpc 
\cite{Jacob,Wischnew}. ICECUBE will monitor the full Galaxy. Within a 
worldwide Supernova Early Warning System \cite{Scholberg}, this 
observation would confirm other records. If several detectors spread 
around the world would measure the signal front with an accuracy of a 
few ms, one possibly might determine the supernova direction by 
triangulation \cite{Weinheimer}.

\section{Acoustic Detection}

Acoustic detection was proposed first in the fifties 
\cite{Ask1}. A particle 
cascade deposits energy into the medium via ionization losses, which is 
converted within a nanosecond into heat. 
The effect is a  steep expansion, generating a bipolar acoustic pulse 
with a width of a microsecond. Transverse to the pencil-like cascade 
(diameter 10 cm)  the radiation is coherent and propagates within a 
disk of about 10 m thickness (the length of the cascade) into the 
medium (see fig.6). 
The frequency peaks at 20\,kHz where
the attenuation length of 
sea water is a few kilometers, compared to a few tens of meters for light. 
Given a large initial signal, huge detection volumes can be achieved. 
The magnitude of the signal is proportional to
%
the heat expansion coefficient  and the cascade energy,
Acoustic pulses from a 200 MeV  beam 
directed into a water tank were measured in 
in 1978 \cite{Sulak}. 
Due to the dependence on the expansion coefficient, 
the signal should disappear at temperatures close to 4$^o$ C.
Warmer water should give a higher 
signal, favoring the Mediterranean with about 14$^o$ C at 4 km depth 
\cite{Butkevitch}. Since the signal also scales with  
the square of the inverse cascade diameter, details of cascade
 simulation are essential for the signal prediction and have been the 
main reason early disagreement between various authors. Following recent 
calculations \cite{Dedenko}, a 10 PeV cascade would generate a signal of 
60\,$\mu$Pa at a distance of 400\,m, which is comparable to the sensitivity of 
the human ear. However, the main challenge of the method is the Ocean 
noise background. 
The signal-to-noise ratio 
may  be improved through coincident detection by many hydrophones close to
 each other as well as at several strings (fig.6). Provided efficient 
noise rejection, acoustic detection might be competetive with optical 
detection at multi-PeV energies. Since recent AGN models suggest several
 events above 1 PeV per year and km$^3$, the method is clearly worth to be 
pursued.

\begin{figure}[htb]
\vspace{9pt}
\centering
\mbox{\epsfig{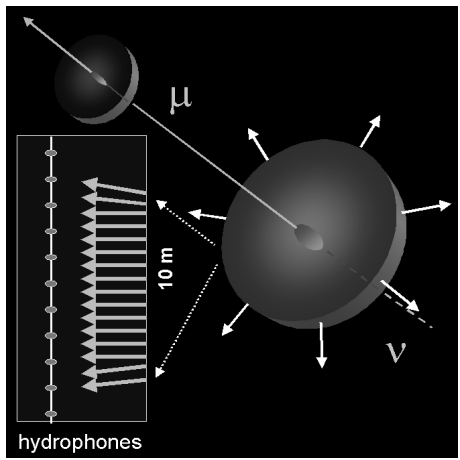}}
\vspace{-0.4cm}
\caption[2]{\small
Principle of acoustic detection
}
\end{figure}

A Russian group has been working on the development of  SADCO, an 
acoustic array at the NESTOR site, which is planned to consist of 
3 strings each instrumented with 128 hydrophones \cite{Butkevitch}.
The  calculated threshold is 6 PeV within a km$^3$ volume. Due to the
priority of building an optical detector, the project is dormant at 
present. 

The same group considers to test an existing sonar array for submarine 
detection close to Kamchatka. The peak sensitivity of the 2400 
hydrophones, however, is only a few hundred Hz, therefore only a 
small fraction of the original signal is captured.  Anyway the 
array might turn out to be useful: Given the large attenuation 
length at these frequencies, it may search for neutrinos with 
energies beyond  $10^{20}$ eV in a volume of 100\,km$^3$ or more. 

First acoustic signals from particle cascades in  open water have
 possibly been identified in Lake Baikal \cite{Domo}. The group operated 
an air shower array on the ice in coincidence with a hydrophone 
installed 5 meters under the ice cover, with the aim to detect the 
acoustic signal generated by the airshower core in water. Indeed a 
series of signals with the proper width of a few hundred 
microseconds have been detected. In order to confirm the nature
 of the pulses, an experiment with better noise reduction and
 higher redundancy will be performed in March 2001.

\section{Radio Detection}

Electromagnetic showers generated by high energy electron neutrino 
interactions emit coherent  Cherenkov radiation \cite{Ask2}. Electrons 
are sweeped into the developing shower, which acquires a negative net 
charge. This charge propagates like a relativistic pancake of 1 cm 
thickness and 10 cm diameter. Each particle emits Cherenkov radiation,
 with the total signal being the resultant of the overlapping 
Cherenkov cones. For wavelengths larger than the cascade diameter, 
 coherence is observed and the signal rises proportional
 to $E^2$, making the method attractive for high energy cascades. The 
bipolar radio pulse has a width of 1-2 ns.  In 
ice, attenuation lengths of several kilometers can be obtained, 
depending on the frequency band and the temperature of the ice. Thus,
 for energies above a few PeV, radio detection in ice promises to be 
superior to optical detection \cite{Price,Frichter}.

\begin{figure}[htb]
\vspace{9pt}
\centering
\mbox{\epsfig{file=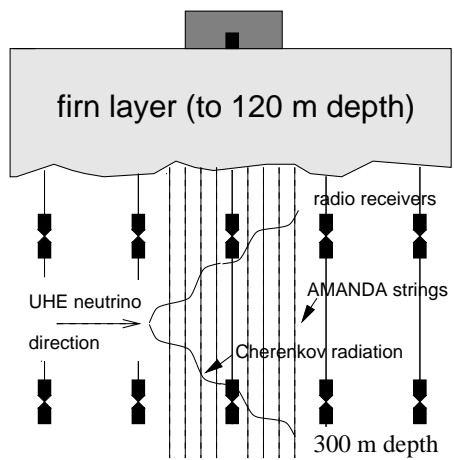,height=6cm}}
\vspace{-0.6cm}
\caption[2]{\small
Schematic view of the RICE array at the South Pole
}
\end{figure}

First studies with respect to noise temperature 
have been performed at the 
russian Vostok station in Antarctica \cite{Provorov}. Meanwhile, a prototype 
Cherenkov radio detector called RICE is operated at the geographical 
South Pole. Twenty receivers and emitters are buried at depths between 
120 and 300 m (see fig.7). Analog signals are red out via  coaxial 
cable, limiting the bandwidth and therefore the fraction of the GHz 
signal arriving at the surface. The use of optical cables would allow 
to go deeper in order to reduce noise from the surface and to get 
better conditions for coincident operation with AMANDA. The data 
analyzed at present 
show that radio sources can be reconstructed 
with about 10 m accuracy. From the non-observation of very large pulses,
first physically relevant upper limits for energies above 100 PeV 
are going to be derived
\cite{Besson}. 

\vspace{-1mm}

\section{Megaton detectors with low threshold}

A future cube kilometer detector will be optimized to energies at or 
above 1 TeV rather than to the low energy range typical for oscillations
 and WIMP search. There is, however, 
much motivation to build  Megaton detectors
 with a lower threshold. 

One proposal suggests an underground 
water-Cherenkov detector like Super-Kamiokande, but  with fiducial 
volume of 650\,ktons \cite{NNN}. The detection threshold of about 7 MeV 
would allow to cover a wide range of physics questions, like solar 
neutrinos, proton decay, search for Supernova bursts, neutrino 
oscillations in the GeV range and search for WIMPs. 
Other schemes proposed employ the ring imaging technique 
\cite{Yps,Eye}, 
with an ultimate size of one Megaton and a similar range of physics
 goals. 

An underwater detector with several Megatons geometrical 
volume and with a threshold of about 5 GeV could bridge the gap 
between the low threshold underground detectors and cube kilometer 
arrays.  It would have a better sensitivity to neutrinos from WIMP 
annihilations than underground detectors and might do astrophysics 
searches in a range complementary to the large arrays. Such an 
detector could be build at the Baikal site or as a sub-detector 
nested in one of the large arrays.

\vspace{-1mm}

\section{Detection at Ultra-High energies}

Radio and acoustic detection may take over from the optical Cherenkov 
methods at energies above 10 PeV.  However, with the exception of the
proposed method to use sonar submarine detection techniques for 
detection of $10^{20}$ eV neutrinos, these methods run out of rate for 
energies in the EeV range ($10^{18}$ eV). 

At higher energies, a large extensive air shower (EAS) array like 
AUGER may seek for horizontal air showers due to neutrino interactions
 deep in the atmosphere. The optimum sensitivity window is between 
$10^{18}$ and $10^{20}$ eV. Given an effective detector mass between 1 and 
20 Giga-tons, the estimated sensitivity is about  
$10^{-7}  E_{\nu}^{-2}$\,cm$^{-2}$\,s$^{-1}$\,sr$^{-1}$\,GeV$^{-1}$. 
 

Heading to higher energies leads to space based detectors monitoring 
larger volumes than visible from any point on the Earth surface. The
EUSO (formerly OWL-Airwatch) project \cite{Owl} 
foresees to launch large mirrors with 
optical detectors to 500 km height. The detector would look down 
upon the atmosphere and search for nitrogen fluorescence signals 
due to neutrino interactions.  The monitored mass would be up to 
10 Tera-tons, with an energy threshold of about $10^{19}$ GeV.

I finally mention an experiment having searched for radio 
emission from extremely-high energy cascades induced by neutrinos or 
cosmic rays in the lunar regolith \cite{Gor}. Using two NASA antennas,
within 12 hours effective data taking an upper limit of
$E_{\nu}^{2} \cdot dN/dE < 10^{-5}$\,cm$^{-2}$\,s$^{-1}$sr$^{-1}$\,GeV at 
$10^{20}$\,eV, has been obtained, close to 
predictions from topological defect models.

\vspace{-1mm}

\section{Conclusions}

After a long period of  development, the first optical underwater/ice
 Cherenkov detectors, BAIKAL and AMANDA,  have detected neutrinos,
 with effective areas in the range of $10^3$ to $10^4$\,m$^2$.  Mediterranean 
detectors are expected to follow soon. However, in order to prove 
most "realistic" models on neutrino production by AGN or GRB, one 
needs kilometer scale detectors. This scale is also suggested by 
the pretention to open a really new window - i.e. to increase the 
sensitivity compared to existing devices by 2-3 orders of magnitude. 
This is a scale which historically nearly inevitably has led to 
unexpected discoveries. Physics as well as economic arguments suggest 
one km$^3$ detector on each hemisphere.

Optical detectors may see only the low energy part of interesting
 phenomena. Therefore further development and funding for acoustic 
as well as radio detection is substantial. These techniques should
 be tested hand in hand with the construction of the big optical 
arrays. 

On the low energy side, Mega-ton detectors with threshold of a few 
GeV or lower could continue the physics program typical for present
 underground detectors with higher sensitivity and complement the 
TeV program of the large arrays by searches in the range below 100 GeV.


\begin{thebibliography}{99}

\bibitem{Ghandi} R.Ghandi, talk at this conference. 

\bibitem{Waxman} E.Waxman, talk at this conference.

\bibitem{Weiler} T.Weiler, talk at this conference.

\bibitem{Gondolo} P.Gondolo, talk at this conference.

\bibitem{oscill} see the talks of H.Sobel, T.Mann, B.Barish,
S.Mikheyev, A.Geiser, P.Lipari and E.Lisi at this conference.

\bibitem{Scholberg} K.Scholberg, talk at this conference.

\bibitem{Zeuthen} Proc.Workshop on
     Large Neutrino Telescopes, ed. C.Spiering,
     Zeuthen, 1998, 
     http://www.ifh.de/nuastro/ publications/ conferences/proc.shtml

\bibitem{Dumand} A.Roberts, Rev.Mod.Phys.64 (1992) 259.

\bibitem{Domo} G.V.Domogatsky, talk at this conference.

\bibitem{Bai1} I.Belolaptikov\,et\,al.,\,Astrop.Phys.7(1997)263.

\bibitem{Bai2} Balkanov,  Astropart.Phys.14 (2000) 61.

\bibitem{Ama1} E.Andres et al., Astropart.Phys. 13 (2000) 1.

\bibitem{Barwick} S.Barwick, talk at this conference.

\bibitem{Nestor} L.Resvanis, talk at this conference.

\bibitem{Antares} L.Thompson, talk at this conference.

\bibitem{NEMO} A.Capone, 26th Int. Cosmic Ray Conf., Salt Lake City 
1999, HE6.3.05.

\bibitem{GHS} T.K.Gaisser, F.Halzen, T.Stanev, Phys.Rev 258 (1996) 355.

\bibitem{LM} J.G.Learned,\,K.Mannheim, High Energy Neutrino Astrophysics, 
to appear in Ann.Rev.Nucl.Phys. (2000).

\bibitem{MPR} K.Mannheim, R.J.Protheroe, J.P.Rachen, Phys.Rev.D (2000),
 astro-ph/9812398.

\bibitem{WB} E.Waxman, J.N.Bahcall, Phys.Rev.D59 (1999) 023002.
\bibitem{Proposal} IceCube: a kilometer-scale neutrino observatory, 
Proposal to NSF, October 1999.

\bibitem{Matthias} M.Leuthold, in \cite{Zeuthen}, p.484. 

\bibitem{Thorn} J.L.Thorn et al., Phys.Rev. D46 (1992) 4846.

\bibitem{Mon} see V.A.Balkanov et al., astro-ph/9906255 and references 
therein.

\bibitem{BEG} L.Bergstroem, J.Edsjo, P.Gondolo, Phys.Rev.D58 (1998) 103519.

\bibitem{Edsjo} J.Edsjo, Amanda Internal Report


\bibitem{Jacob} F.Halzen,J.Jacobsen, Phys.Rev.D49 (1994) 1758.

\bibitem{Wischnew} R.Wischnewski et al., Proc. 26th Int. Cosmic 
Ray Conf.,\,Salt\,Lake\,City\,1999,\,HE.4.2.07.

\bibitem{Weinheimer} L.K\"opke and C. Weinheimer, Amanda Int. Report, 
April 2000. See also J.F.Beacom and P.Vogel, astro-ph/9811350v2.

\bibitem{Ask1} G.A.Askaryan,\,Sov.\,Journ.\,Atom.\,Energy\,3
(1957)\,921, Nucl.\,Instr.\,Meth.164\,(1979)\,267.

\bibitem{Sulak} L.Sulak et al., Nucl.Instr.Meth. 161 (1979) 203. 
See also J.G.Learned, Phys.Rev.D 19 (1979) 3293.

\bibitem{Butkevitch} A.V.Butkevitch et al., Proc.Int. School an Part. 
and Cosmology, Baksan Valley, World Scientific (1996) 306.

\bibitem{Dedenko} L.G.Dedenko et al., Proc. 25th Int. Cosmic Ray Conf.,
Durban 1997, HE.4.1.27.

\bibitem{Ask2} G.Askaryan, Sov.Phys.JETP 14 (1962) 441.

\bibitem{Price} P.B.Price, Astropart.Phys 5 (1996) 43.

\bibitem{Frichter} G.M.Frichter, J.P.Ralston, D.W.McKay , 
Phys.Rev. D53 (1996) 1684.

\bibitem{Provorov} A.Provorov,\,I.Zhelesnykh,\,Astrop.Phys.4 (1995)\,55.

\bibitem{Besson} D.Besson et al., to be published.

\bibitem{NNN} see e.g. C.K.Jung, hep-ex/005046.

\bibitem{Yps} P.Antonioli et al., Nucl.Instr.Meth. A433 (1999) 104. 

\bibitem{Eye} J.G.Learned, in \cite{Zeuthen}, p.466.

\bibitem{AUGER} Auger Technical Design Report, 
Fermilab-Pub-96-024 (1996).

\bibitem{Owl} L.Scarsi et.al, Proc. 26th Int.Cosmic Ray Conf., 
Salt Lake City 1999, HE 6.1.07.

\bibitem{Gor} P.W.Gorham et al., astro-ph/9906504.

\end{thebibliography}
\end{document}